\documentclass[conference, hidelinks, 10pt]{IEEEtran}
\IEEEoverridecommandlockouts

\usepackage{amsmath,amssymb,amsfonts}
\usepackage{algorithmic}
\usepackage{graphicx}
\usepackage{textcomp}
\usepackage[table,xcdraw]{xcolor}
\usepackage{orcidlink}
\usepackage[farskip=0pt]{subfig}
\usepackage{lipsum}
\usepackage[acronym]{glossaries}
\usepackage[detect-all=true,per-mode=repeated-symbol,per-symbol =/]{siunitx}
\usepackage[capitalise]{cleveref}
\usepackage{tablefootnote}
\usepackage{threeparttable}
\usepackage{booktabs}
\usepackage{adjustbox}
\usepackage{multirow}
\usepackage{ragged2e}
\usepackage{microtype}
\usepackage{paralist}
\usepackage{float}
\usepackage{placeins}
\usepackage[backend=biber,style=numeric-comp,sorting=none, maxcitenames=1, mincitenames=1, maxbibnames=1]{biblatex}
\usepackage{geometry}
\usepackage{makecell}
\usepackage{soul}
\usepackage{pifont}

\definecolor{ieee-bright-dblue-100}{rgb}{0.0, 0.3828, 0.6055}
\definecolor{ieee-bright-dblue-80}{rgb}{0.0, 0.4883, 0.6797}
\definecolor{ieee-bright-dblue-60}{rgb}{0.3633, 0.6094, 0.7617}
\definecolor{ieee-bright-dblue-40}{rgb}{0.5898, 0.7383, 0.8398}
\definecolor{ieee-bright-dblue-20}{rgb}{0.8906, 0.8984, 0.9219}
\definecolor{ieee-bright-red-100}{rgb}{0.7266, 0.0469, 0.1836}
\definecolor{ieee-bright-red-80}{rgb}{0.832, 0.3164, 0.3281}
\definecolor{ieee-bright-red-60}{rgb}{0.8906, 0.4922, 0.4805}
\definecolor{ieee-bright-red-40}{rgb}{0.9336, 0.6562, 0.6406}
\definecolor{ieee-bright-red-20}{rgb}{0.9688, 0.8203, 0.8125}
\definecolor{ieee-bright-orange-100}{rgb}{0.9961, 0.6367, 0.0}
\definecolor{ieee-bright-orange-80}{rgb}{0.9844, 0.6953, 0.3125}
\definecolor{ieee-bright-orange-60}{rgb}{0.9883, 0.7695, 0.4844}
\definecolor{ieee-bright-orange-40}{rgb}{0.9922, 0.8359, 0.6562}
\definecolor{ieee-bright-orange-20}{rgb}{0.9961, 0.9219, 0.8164}
\definecolor{ieee-bright-yellow-100}{rgb}{0.9961, 0.8164, 0.0}
\definecolor{ieee-bright-yellow-80}{rgb}{0.9961, 0.8477, 0.2148}
\definecolor{ieee-bright-yellow-60}{rgb}{0.9961, 0.875, 0.4492}
\definecolor{ieee-bright-yellow-40}{rgb}{0.9961, 0.9062, 0.6328}
\definecolor{ieee-bright-yellow-20}{rgb}{0.9961, 0.9531, 0.8125}
\definecolor{ieee-bright-lgreen-100}{rgb}{0.4688, 0.7422, 0.125}
\definecolor{ieee-bright-lgreen-80}{rgb}{0.5742, 0.7852, 0.332}
\definecolor{ieee-bright-lgreen-60}{rgb}{0.6875, 0.8398, 0.5039}
\definecolor{ieee-bright-lgreen-40}{rgb}{0.793, 0.8906, 0.6641}
\definecolor{ieee-bright-lgreen-20}{rgb}{0.8945, 0.9414, 0.8281}
\definecolor{ieee-bright-dgreen-100}{rgb}{0.0, 0.5156, 0.2383}
\definecolor{ieee-bright-dgreen-80}{rgb}{0.1641, 0.6055, 0.3867}
\definecolor{ieee-bright-dgreen-60}{rgb}{0.3906, 0.6953, 0.5234}
\definecolor{ieee-bright-dgreen-40}{rgb}{0.6094, 0.8008, 0.6719}
\definecolor{ieee-bright-dgreen-20}{rgb}{0.8047, 0.8945, 0.8359}
\definecolor{ieee-bright-purple-100}{rgb}{0.5938, 0.1133, 0.5898}
\definecolor{ieee-bright-purple-80}{rgb}{0.6992, 0.3281, 0.668}
\definecolor{ieee-bright-purple-60}{rgb}{0.7812, 0.4961, 0.7461}
\definecolor{ieee-bright-purple-40}{rgb}{0.8555, 0.6602, 0.8281}
\definecolor{ieee-bright-purple-20}{rgb}{0.9219, 0.8281, 0.9023}
\definecolor{ieee-bright-lblue-100}{rgb}{0.0, 0.6094, 0.6484}
\definecolor{ieee-bright-lblue-80}{rgb}{0.0, 0.6797, 0.7188}
\definecolor{ieee-bright-lblue-60}{rgb}{0.2109, 0.75, 0.7812}
\definecolor{ieee-bright-lblue-40}{rgb}{0.5469, 0.8242, 0.8438}
\definecolor{ieee-bright-lblue-20}{rgb}{0.7695, 0.918, 0.9219}
\definecolor{ieee-bright-cyan-100}{rgb}{0.0, 0.707, 0.8828}
\definecolor{ieee-bright-cyan-80}{rgb}{0.0, 0.7227, 0.9453}
\definecolor{ieee-bright-cyan-60}{rgb}{0.2656, 0.7812, 0.957}
\definecolor{ieee-bright-cyan-40}{rgb}{0.5547, 0.8438, 0.9688}
\definecolor{ieee-bright-cyan-20}{rgb}{0.7773, 0.9141, 0.9805}
\definecolor{ieee-bright-white-100}{rgb}{0.9961, 0.9961, 0.9961}
\definecolor{ieee-bright-white-80}{rgb}{0.9961, 0.9961, 0.9961}
\definecolor{ieee-bright-white-60}{rgb}{0.9961, 0.9961, 0.9961}
\definecolor{ieee-bright-white-40}{rgb}{0.9961, 0.9961, 0.9961}
\definecolor{ieee-bright-white-20}{rgb}{0.9961, 0.9961, 0.9961}
\definecolor{ieee-dark-red-100}{rgb}{0.5234, 0.1211, 0.2539}
\definecolor{ieee-dark-red-80}{rgb}{0.6445, 0.2812, 0.3828}
\definecolor{ieee-dark-red-60}{rgb}{0.7422, 0.4727, 0.5234}
\definecolor{ieee-dark-red-40}{rgb}{0.832, 0.6445, 0.6758}
\definecolor{ieee-dark-red-20}{rgb}{0.918, 0.8203, 0.832}
\definecolor{ieee-dark-orange-100}{rgb}{0.9062, 0.4648, 0.1328}
\definecolor{ieee-dark-orange-80}{rgb}{0.9648, 0.5664, 0.3164}
\definecolor{ieee-dark-orange-60}{rgb}{0.9766, 0.6758, 0.4805}
\definecolor{ieee-dark-orange-40}{rgb}{0.9844, 0.7773, 0.6523}
\definecolor{ieee-dark-orange-20}{rgb}{0.9922, 0.8789, 0.8125}
\definecolor{ieee-dark-yellow-100}{rgb}{0.9961, 0.7773, 0.1719}
\definecolor{ieee-dark-yellow-80}{rgb}{0.9961, 0.8086, 0.375}
\definecolor{ieee-dark-yellow-60}{rgb}{0.9961, 0.875, 0.4492}
\definecolor{ieee-dark-yellow-40}{rgb}{0.9961, 0.8984, 0.6875}
\definecolor{ieee-dark-yellow-20}{rgb}{0.9961, 0.9453, 0.8438}
\definecolor{ieee-dark-lgreen-100}{rgb}{0.3945, 0.5508, 0.0938}
\definecolor{ieee-dark-lgreen-80}{rgb}{0.5078, 0.6289, 0.293}
\definecolor{ieee-dark-lgreen-60}{rgb}{0.6367, 0.7188, 0.4688}
\definecolor{ieee-dark-lgreen-40}{rgb}{0.7539, 0.8047, 0.6367}
\definecolor{ieee-dark-lgreen-20}{rgb}{0.875, 0.9023, 0.8125}
\definecolor{ieee-dark-dgreen-100}{rgb}{0.0, 0.3867, 0.2539}
\definecolor{ieee-dark-dgreen-80}{rgb}{0.1836, 0.5, 0.3906}
\definecolor{ieee-dark-dgreen-60}{rgb}{0.3984, 0.6172, 0.5273}
\definecolor{ieee-dark-dgreen-40}{rgb}{0.5938, 0.7422, 0.6758}
\definecolor{ieee-dark-dgreen-20}{rgb}{0.793, 0.8711, 0.8359}
\definecolor{ieee-dark-purple-100}{rgb}{0.4648, 0.1445, 0.5117}
\definecolor{ieee-dark-purple-80}{rgb}{0.5898, 0.3242, 0.6016}
\definecolor{ieee-dark-purple-60}{rgb}{0.6914, 0.4883, 0.6953}
\definecolor{ieee-dark-purple-40}{rgb}{0.7969, 0.6523, 0.793}
\definecolor{ieee-dark-purple-20}{rgb}{0.8945, 0.8203, 0.8945}
\definecolor{ieee-dark-cyan-100}{rgb}{0.0, 0.4492, 0.4648}
\definecolor{ieee-dark-cyan-80}{rgb}{0.0, 0.5469, 0.5664}
\definecolor{ieee-dark-cyan-60}{rgb}{0.3047, 0.6602, 0.668}
\definecolor{ieee-dark-cyan-40}{rgb}{0.5586, 0.7695, 0.7734}
\definecolor{ieee-dark-cyan-20}{rgb}{0.7734, 0.8789, 0.8789}
\definecolor{ieee-dark-dblue-100}{rgb}{0.0, 0.1562, 0.332}
\definecolor{ieee-dark-dblue-80}{rgb}{0.1797, 0.3008, 0.4609}
\definecolor{ieee-dark-dblue-60}{rgb}{0.3828, 0.4609, 0.5859}
\definecolor{ieee-dark-dblue-40}{rgb}{0.5781, 0.6289, 0.7188}
\definecolor{ieee-dark-dblue-20}{rgb}{0.7852, 0.8047, 0.8555}
\definecolor{ieee-dark-grey-100}{rgb}{0.457, 0.4688, 0.4805}
\definecolor{ieee-dark-grey-80}{rgb}{0.5625, 0.5625, 0.5742}
\definecolor{ieee-dark-grey-60}{rgb}{0.6641, 0.6641, 0.6758}
\definecolor{ieee-dark-grey-40}{rgb}{0.7734, 0.7695, 0.7773}
\definecolor{ieee-dark-grey-20}{rgb}{0.8789, 0.8828, 0.8828}
\definecolor{ieee-dark-black-100}{rgb}{0.0, 0.0, 0.0}
\definecolor{ieee-dark-black-80}{rgb}{0.3438, 0.3477, 0.3555}
\definecolor{ieee-dark-black-60}{rgb}{0.5, 0.5078, 0.5195}
\definecolor{ieee-dark-black-40}{rgb}{0.6523, 0.6602, 0.6719}
\definecolor{ieee-dark-black-20}{rgb}{0.8164, 0.8242, 0.8281}

\renewbibmacro{in:}{, }

\DeclareFieldFormat{year}{\addcomma\space#1}
\DeclareFieldFormat[inproceedings]{booktitle}{\textit{#1}\addcomma\space}
    
\renewbibmacro*{issue+date}{%
    \printfield{year}%
    \newunit}

\AtEveryBibitem{%
    \clearfield{title}\clearfield{publisher}\clearfield{location}%
    \clearfield{volume}\clearfield{number}\clearfield{pages}%
    \clearfield{series}\clearfield{note}%
}

\defbibenvironment{bibliography}{}{}
  {\small\noindent\printtext[labelnumberwidth]{%
     \printfield{labelprefix}%
     \printfield{labelnumber}}%
   \addhighpenspace}
\DeclareSIUnit{\x}{\!\ensuremath{\times}}
\DeclareSIUnit\bit{b}
\DeclareSIUnit\flop{FLOP}
\DeclareSIUnit\dash{\text{-}}
\DeclareSIUnit\gateeq{GE}
\newcommand{\basilisk}{Basilisk}
\newcommand{\cheshire}{Cheshire}
\newcommand{\riscv}{\mbox{RISC-V}}
\newacronym[longplural={systems-on-chip}]{soc}{SoC}{system-on-chip}
\newacronym{oseda}{OSEDA}{open-source electronic design automation}
\newacronym{soa}{SoA}{state-of-the-art}
\newacronym{eda}{EDA}{electronic design automation}
\newacronym{orfs}{ORFS}{OpenROAD flow scripts}
\newacronym{pnr}{P\&R}{place and route}
\newacronym{qor}{QoR}{quality of results}
\newacronym{sv}{SV}{SystemVerilog}
\newacronym{ip}{IP}{intellectual propertie}
\newacronym{rtl}{RTL}{register transfer level}
\newacronym{hdl}{HDL}{hardware description language}
\newacronym{pdk}{PDK}{process design kit}
\newacronym{drc}{DRC}{design rule check}
\newacronym{axi}{AXI}{advanced eXtensible interface}
\newacronym{axi4}{AXI4}{advanced eXtensible interface 4}
\newacronym{llc}{LLC}{last-level cache}
\newacronym{c2c}{C2C}{chip-to-chip}
\newacronym{gpt}{GPT}{Globally Unique Identifier Partition Table}
\newacronym{gemm}{GEMM}{general matrix multiply}
\newacronym{ll}{LL}{levels of logic}

\newcommand{\x}{$\times$}

\newcommand{\cmark}{\ding{51}}%
\newcommand{\xmark}{\ding{55}}%

\title{\fontsize{16pt}{19pt}\selectfont\bf
Basilisk: A 34 mm² End-to-End Open-Source\\
64-bit Linux-Capable RISC-V SoC in 130nm BiCMOS
\vspace{-0.3cm}
}

\begin{document}

\ifdefined\blind
    \author{%
    \vspace{0.1cm} %
    \textit{Authors omitted for blind review}
    \vspace{0.1cm} %
    }
\else
\author{
    \IEEEauthorblockN{%
    Philippe Sauter\orcidlink{0009-0001-6504-8086}\IEEEauthorrefmark{1}\IEEEauthorrefmark{10}, %
    Thomas Benz\orcidlink{0000-0002-0326-9676}\IEEEauthorrefmark{1}\IEEEauthorrefmark{10}, %
    Paul Scheffler\orcidlink{0000-0003-4230-1381}\IEEEauthorrefmark{1}\IEEEauthorrefmark{10}, %
    Martin Povišer\orcidlink{0000-0002-0012-5319}, %
    Frank K. G\"urkaynak\orcidlink{0000-0002-8476-554X}\IEEEauthorrefmark{1}, %
    Luca Benini\orcidlink{0000-0001-8068-3806}\IEEEauthorrefmark{1}\IEEEauthorrefmark{2}%
    }
    \thanks{%
        \IEEEauthorrefmark{10} All authors contributed equally to this research.
    }
    \IEEEauthorblockA{
        \textasteriskcentered~\textit{%
        ETH Zurich}, Switzerland \hspace{1.5cm}
        \textdagger~\textit{%
        University of Bologna}, Italy \\
        }
    \vspace{-0.75cm}
    }
\fi

\maketitle

\begin{abstract}

End-to-end \gls{oseda} enables a collaborative approach to chip design conducive to supply chain diversification and zero-trust step-by-step design verification.
However, existing end-to-end \gls{oseda} flows have mostly been demonstrated on small designs and have not yet enabled large, industry-grade chips such as Linux-capable \glspl{soc}.
This work presents {\basilisk}, the largest end-to-end open-source \gls{soc} to date. 
{\basilisk}'s \SI{34}{\milli\meter^2}, \SI{2.7}{\mega\gateeq} design features a 64-bit Linux-capable RISC-V core, a lightweight \SI{124}{\mega\byte\per\second} DRAM controller, and extensive IO, including a USB 1.1 host, a video output, and a fully digital \SI{62}{\mega\bit\per\second} \gls{c2c} link.
We implement {\basilisk} in IHP's open \SI{130}{\nano\meter} BiCMOS technology, significantly improving on the \gls{soa} \gls{oseda} flow. 
Our enhancements of the Yosys-based synthesis flow improve design timing and area by 2.3\x and 1.6\x, respectively, while consuming significantly less system resources. By tuning OpenROAD \gls{pnr} to our design and technology, we decrease the die size by 12\%.
The fabricated {\basilisk} chip reaches \SI{62}{\mega\hertz} at its nominal \SI{1.2}{\volt} core voltage and up to \SI{102}{\mega\hertz} at \SI{1.64}{\volt}.
It achieves a peak energy efficiency of \SI{18.9}{DP~\mega FLOP\per\second\per\watt} at \SI{0.88}{\volt}.

\end{abstract}

\glsresetall

\section{Introduction}

Many recently funded \emph{Chips Acts} ~\cite{2024europeanchipsact, 2024chipsandscienceact, 2024indiainjects} aim at creating robust domestic silicon value chains, significantly increasing the number of chip designers, and reducing the cost and friction to access silicon manufacturing.
One promising approach to realize these goals is \emph{open-source silicon}:
by sharing their designs under permissive open-source licenses, chip designers can freely collaborate and build on each other's work.
In fact, an \emph{end-to-end} open design flow, shown in \cref{fig:flow}, would empower \emph{anyone} to independently verify an entire chip step by step, enabling its safe use without requiring blind trust in the supply chain stakeholders.
Thanks to a recent flourishing of open-source designs~\cite{ottaviano2023cheshire, zaruba2019ariane}, automation tools~\cite{wolf2013yosys, ajayi2019openroad}, and \glspl{pdk}~\cite{herman2024versatility}, end-to-end \gls{oseda} flows~\cite{orfs-git, ghazy2020openlane} already exist and produce viable results on smaller designs~\cite{henkes2024evaluating, khan2023ghazi, khan2021ibtida, zhu2022greenrio}.
However, these open flows still lag behind their commercial closed-source counterparts and have not yet enabled designers to implement and optimize large, industry-grade designs such as application-class Linux-capable \glspl{soc}.

In this work, we present first silicon measurements of 
{\basilisk}\footnote{{\basilisk} source: \url{https://github.com/pulp-platform/cheshire-ihp130-o}}, to the best of our knowledge, the largest end-to-end open-source \gls{soc} to date. 
{\basilisk} is a Linux-capable 64-bit RISC-V \gls{soc} with a hierarchical interconnect, a lightweight \SI{124}{\mega\byte\per\second} DRAM controller, and extensive IOs:
UART, I2C, QSPI, a VGA video output, a four-port USB 1.1 host, and a fully digital \SI{62}{\mega\bit\per\second} \gls{c2c} link.
We implemented {\basilisk}'s open \SI{2.7}{\mega\gateeq} %
design, fully based on open-source \glspl{ip}, in IHP's open \SI{130}{\nano\meter} BiCMOS \gls{pdk} using \gls{oseda} tools, significantly improving on the \gls{soa} open flow in the process.
We developed a capable open \gls{sv} frontend and optimized logic synthesis using \emph{Yosys}, improving cell area by 1.6\x~and timing by 2.3\x~while reducing synthesis runtime by 2.5\x and its memory usage by 4.8\x.
We further tuned \gls{pnr} in \emph{OpenROAD} to our design, increasing core utilization by \SI{10}{\percent}.
The fabricated \SI{34}{\milli\meter^2} chip reaches a clock frequency of \SI{62}{\mega\hertz}, corresponding to 51 \gls{ll}~\footnote{Number of gates in longest path}, at the nominal \SI{1.2}{\volt} core voltage and up to \SI{102}{\mega\hertz} at \SI{1.64}{\volt}.
On FP64 \gls{gemm}, we measure a peak energy efficiency of~\SI{18.9}{\mega\flop\per\second\per\watt} at the minimum operational core voltage of \SI{0.88}{\volt} at \SI{10}{\mega\hertz}.

\section{Architecture}

\begin{figure*}[t!]
    \begin{minipage}{0.64\linewidth}
        \centering
        \includegraphics[width=\linewidth]{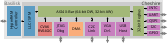}
        \vspace{-0.5cm}
        \captionof{figure}{\label{fig:arch}Top-level architecture of {\basilisk} built on the {\cheshire}~\cite{ottaviano2023cheshire} \gls{soc} platform.}
        \vspace{0.4cm}
        \begin{minipage}{0.48\linewidth}
            \centering
            \vspace{-0.23cm}
            \includegraphics[width=\linewidth]{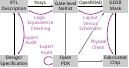}
            \vspace{-0.17cm}
            \captionof{figure}{\label{fig:flow} End-to-end open-source chip design flow and verification of each step.}
            \vspace{0.2cm}
        \end{minipage}
            \hspace{0.05cm}
        \begin{minipage}{0.48\linewidth}
            \centering
            \includegraphics[width=\linewidth]{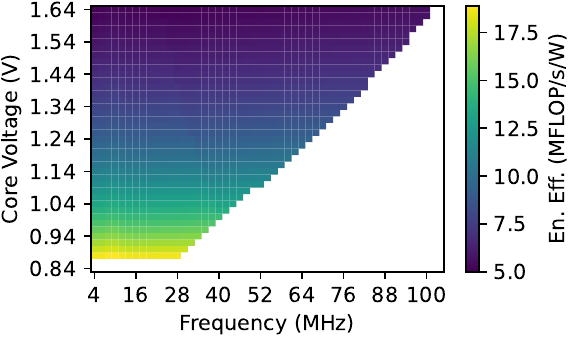}
            \vspace{-0.5cm}
            \captionof{figure}{\label{fig:shmoo}Shmoo plot at \SI{25}{\celsius} annotated with FP64 GEMM energy efficiencies.}
            \vspace{0.4cm}
        \end{minipage}
        \resizebox{\columnwidth}{!}{%
    \centering
        \renewcommand*{\arraystretch}{0.95}
        \begin{threeparttable}
            \begin{tabular}{@{}lcc|ccccc@{}}
            
                \toprule
            
                &
                \textit{\textbf{Basilisk}}~[ours] &
                \textit{MLEM}~\tnote{a}~\cite{sauter2025crocendtoendopensourceextensible} &
                \textit{HEP}~\cite{henkes2024evaluating} &
                \textit{Ghazi}~\cite{khan2023ghazi} &
                \textit{Ibtida}~\cite{khan2021ibtida} &
                \textit{GreenRio}~\cite{zhu2022greenrio} \\

                \midrule

                \textit{Chip Area} [\si{\micro\metre\squared}] &
                34.4 &
                5.0 &
                13.7 &
                6.9 &
                2.6 &
                2.0 \\

                \textit{Chip Complexity} [\si{\kilo\gateeq}] &
                2700 &
                350 &
                - &
                $\approx$500 &
                340 &
                - \\

                \textit{Std. Cell Complexity} [\si{\kilo\gateeq}] &
                1140 &
                102 &
                230~\tnote{b} &
                $\approx$200 &
                94 &
                53 \\

                \textit{Synthesis Speed} [\si{\MHz}] &
                77 &
                80 &
                25 &
                12.5 &
                - &
                80 \\

                \textit{Silicon Speed} [\si{\MHz}] &
                62 &
                80 &
                25 &
                12.5 &
                - &
                80 \\

                \textit{Logic levels (LL)~\tnote{c}} &
                51 &
                58 &
                - &
                - &
                - &
                - \\

                \textit{Synth. Time} [\si{{\minute}}] &
                130~\tnote{d} / 110~\tnote{b}~\tnote{e} &
                6.6~\tnote{d} / 65~\tnote{b}~\tnote{e} &
                - &
                - &
                - &
                6.2 / 120~\tnote{e} \\

                \textit{Impl. Time} [\si{{\hour}}] &
                24~\tnote{d} / 21~\tnote{e} &
                0.9~\tnote{d} / 8.8~\tnote{e} &
                - &
                - &
                - &
                2.0 / 38~\tnote{e} \\

                \textit{Open PDK} &
                IHP130 &
                IHP130 &
                \textcolor{ieee-dark-red-100}{\xmark} &
                SKY130 &
                SKY130 &
                SKY130 \\

                \textit{Application-Class} &
                \textcolor{ieee-dark-dgreen-100}{\cmark} &
                \textcolor{ieee-dark-red-100}{\xmark} &
                \textcolor{ieee-dark-red-100}{\xmark} &
                \textcolor{ieee-dark-red-100}{\xmark} &
                \textcolor{ieee-dark-red-100}{\xmark} &
                \textcolor{ieee-dark-red-100}{\xmark} \\

                \bottomrule
                
            \end{tabular}

            \begin{tablenotes}[para, flushleft]
                \item[a] Simple microcontroller reusing {\basilisk}'s flow
                \item[b] Assuming 50\% placement density
                \item[c] Number of gates in longest path
                \item[d] \SI{2.5}{\GHz} Xeon E5-2670
                \item[e] Normalized with standard cell complexity in \si{\mega\gateeq}

            \end{tablenotes}
        \end{threeparttable}
}

        \vspace{-0.0cm}
        \captionof{table}{\label{tab:soa}Comparison to other open \SI{130}{\nano\meter} designs  implemented with open flows.}
        \vspace{0.0cm}
    \end{minipage}
    \hspace{0.05cm}
    \begin{minipage}{0.343\linewidth}
        \centering
        \includegraphics[width=\linewidth]{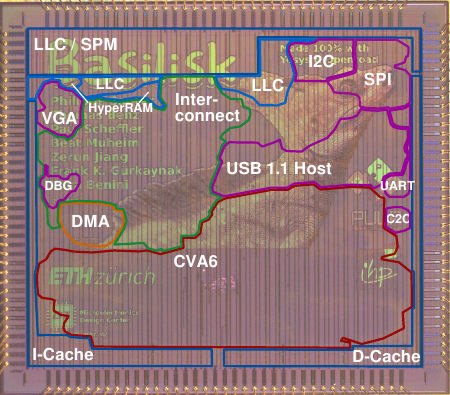}
        \vspace{-0.5cm}
        \captionof{figure}{\label{fig:die}{\basilisk} die shot and floorplan.}
        \vspace{0.1cm}
        \includegraphics[width=\linewidth]{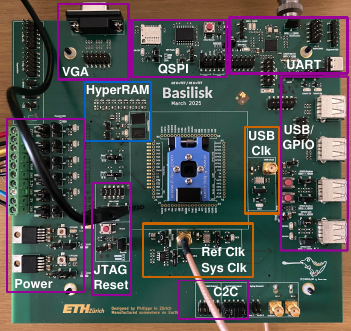}
        \vspace{-0.4cm}
        \captionof{figure}{\label{fig:board}{\basilisk} evaluation board.}
        \vspace{0.0cm}
    \end{minipage}

    \vspace{-0.1cm}
    
\end{figure*} %

\cref{fig:arch} shows Basilisk's top-level architecture, which is based on our open-source, permissively licensed \emph{Cheshire}~\cite{ottaviano2023cheshire} \gls{soc} platform and uses the RV64GC-compliant \emph{CVA6}~\cite{zaruba2019ariane} core.
Basilisk integrates all hardware \glspl{ip} needed to run Linux, including RISC-V-compliant interrupt controllers and a DRAM interface.
To balance performance and area, Basilisk features a two-stage interconnect: high-throughput components attach to a fully connected 64-bit \glsunset{axi4}\gls{axi4}~\cite{kurth2021open} crossbar, while low-throughput peripherals attach to a lightweight Regbus~\cite{register-interface-git} demultiplexer. %

Basilisk's fully digital HyperRAM controller supports two DRAM chips and transfer speeds of up to \SI{124}{\mega\byte\per\second}. 
It connects to the \gls{axi4} crossbar through a four-way, \SI{473}{\mega\byte\per\second}, \SI{64}{\kibi\byte} \gls{llc},
which can have each of its ways dynamically configured as a scratchpad memory.
The CVA6 core is configured with four-way, \SI{16}{\kibi\byte} L1 instruction and data caches.%

Basilisk provides a rich set of peripherals.
In addition to I2C, QSPI, and UART for serial communication, it includes a fully digital four-port \SI{12}{\mega\bit\per\second} USB 1.1 host controller and a VGA video output supporting \emph{XGA} at \SI{60}{\hertz}.
Each USB port is multiplexed with GPIOs, providing a software-controlled IO bus up to 8 bits wide.
A JTAG test access point connected to a \riscv~debug module enables live debugging of the CVA6 core and full memory access.
A fully digital duplex \SI{62}{\mega\bit\per\second} \gls{c2c} link enables inter-chip communication through direct memory accesses.
A high-efficiency \SI{473}{\mega\byte\per\second}, asynchronous DMA engine~\cite{benz2023high} capable of 2D transfers can relieve CVA6 of data movement tasks.
Basilisk's peripherals are designed to be Linux-compatible, with many already having working drivers.

\section{Open-Source Implementation}

Industry-grade \gls{sv} code presents a significant challenge for \gls{oseda} toolchains.
Existing approaches like \emph{SV2V}~\cite{sv2v-git} convert \gls{sv} to simpler Verilog before elaboration.
However, this bloats and obfuscates the hardware description, increasing synthesis runtime and memory usage while degrading the \gls{qor}~\cite{sauter2024insights}.
To address this shortcoming, we developed \emph{Yosys-Slang}\footnote{Yosys-slang source: \url{https://github.com/povik/yosys-slang}}, an \gls{sv} frontend for Yosys based on the leading open \gls{sv} compiler \emph{Slang}~\cite{slang-git}. 
Unlike existing solutions, Yosys-Slang can directly elaborate industry-grade \gls{sv} code and thus preserve design intent.
This results in smaller netlists, lower memory usage, shorter runtime, better debuggability, and improved \gls{qor}.

We improved logic synthesis in Yosys. We added a specialized optimization mapping aligned shift operations to multiplexers and tuned the synthesis flow for our technology and design. 
We leveraged \emph{lazy man's synthesis}~\cite{lazy-synthesis} to build a high-effort logic optimization script. %
Compared to a baseline synthesis flow derived from~\cite{orfs-git}, our \gls{sv} frontend and synthesis improvements reduce standard cell areafrom 1.8 to \SI{1.1}{\mega\gateeq} (1.6\x) and increase post-synthesis clock frequency from 33 to \SI{77}{\mega\hertz} (2.3\x). They simultaneously reduce synthesis runtime from 5.4 to \SI{1.7}{\hour} (3.2\x) and its memory footprint from 35 to \SI{7.4}{\giga\byte} (4.8\x).

Finally, starting from the \gls{orfs}~\cite{orfs-git}, we tuned OpenROAD's flow and hyperparameters in the global placement and routing stages to our design by better balancing routing and timing goals. 
We strategically added small routing blockages, especially around power grid connections, to guide local routing behavior.
This significantly reduced routing congestion in dense modules like the boot ROM, allowing us to reduce the die area %
from 39 to \SI{34}{\milli\meter^2} (-\SI{12}{\percent}). %

\section{Silicon Results}

\cref{fig:die} shows the fabricated {\basilisk} chip annotated with its final floorplan; the core area is dominated by the CVA6 core (\SI{45.6}{\percent}) and \gls{llc} (\SI{12.5}{\percent}).
To characterize {\basilisk}, we created the evaluation board shown in \cref{fig:board}, which breaks out its peripheral IO and provides the necessary circuitry for standalone operation.
We use \emph{DUTCTL}~\cite{benz2024dutctl} to orchestrate on-chip workload execution with the external power supplies and clock sources.

We ran a 48\x 48 FP64 \gls{gemm} on {\basilisk} at \SI{25}{\celsius} while sweeping the core voltage and frequency, resulting in the annotated Shmoo plot shown in \cref{fig:shmoo}.
At the nominal \SI{1.2}{\volt} voltage, {\basilisk} reaches a clock frequency of up to \SI{62}{\mega\hertz}. 
At \SI{1.64}{\volt}, we reach {\basilisk}'s peak frequency of \SI{102}{\mega\hertz}.
{\basilisk} reaches a competitive technology-normalized speed of 51 \gls{ll} compared to another commercially implemented {\cheshire}-based design~\cite{ottaviano2023cheshire} at 46 \gls{ll}.
At the minimum operational core voltage of \SI{0.88}{\volt}, we observe the peak energy efficiency of~\SI{18.9}{\mega\flop\per\second\per\watt}.
Among ten packaged chips, we observe an average leakage power of \SI{437}{\micro\watt}.

\cref{tab:soa} compares {\basilisk} to four other open \SI{130}{\nano\metre} chips taped out using open design flows.
{\basilisk} is the only application-class \gls{soc} among them, exceeding the next-largest design's standard cell complexity by 4.8\x~while maintaining competitive clock speeds.
At \SI{110}{\minute\per\mega\gateeq} and \SI{21}{\hour\per\mega\gateeq}, {\basilisk}'s complexity-normalized synthesis and \gls{pnr} runtimes outperforms \emph{GreenRio}'s~\cite{zhu2022greenrio}. In fact, these runtimes are only bested by those of MLEM~\cite{sauter2025crocendtoendopensourceextensible}, a simple microcontroller design \emph{reusing} {\basilisk}'s tuned \gls{oseda} flow; this demonstrates the portability of our implementation flow improvements to other designs.

\section*{Acknowledgments}
\noindent
We thank
J. Schoenleber,
A. Ottaviano,
N. Wistoff,
A. Prasad,
C. Koenig,
R. Balas,
C. Reinwardt,
N. Narr,
and T. Senti.
We also thank IHP for their generous support.
Silicon manufacturing was sponsored by German BMBF project FMD-QNC (16ME0831).

\printbibliography

\end{document}